\documentclass[a4,draftcls,12pt,onecolumn]{IEEEtran}
\usepackage{cite}
\usepackage{epsfig,graphics,mathrsfs,amssymb,amsmath,bm,color,yfonts,txfonts,multirow,multicol,slashbox,dblfloatfix}
\usepackage{subfigure}
\usepackage{algorithm2e}
\usepackage[noend]{algpseudocode}

\begin{document}

\title{Shared Spectrum Access Communications: A Neutral Host Micro Operator Approach}
\author{Mirza Golam Kibria, \textit{Member, IEEE,} Gabriel Porto Villardi, \textit{Senior Member, IEEE,} Kien Nguyen, \textit{Senior Member, IEEE,} Wei-Shun Liao, \textit{Member, IEEE,} Kentaro Ishizu and Fumihide Kojima, \textit{Member, IEEE}
 \thanks{The authors are with Wireless Systems Laboratory,  Wireless Networks Research Center, National Institute of Information and Communications Technology (NICT), 3-4 Hikarino-oka, Yokosuka, Japan 239-0847 (e-mail: mirza.kibria@nict.go.jp, gpvillardi@nict.go.jp, kienng@nict.go.jp, wsliao@nict.go.jp, ishidu@nict.go.jp, f-kojima@nict.go.jp). }
}
\maketitle

\begin{abstract}

In this paper, we conceive an advanced neutral host micro operator (NH-$\mu$O) network approach providing venues with services tailored to their specialized/specific requirements and/or local context related services that the mobile network operators (MNOs) are poorly-suited to providing it, as well as mobile broadband experience to the users from MNOs in a venue where only a single infrastructure is mandated under shared spectrum access framework.
A radio access network slicing concept is conceived to support and optimize both the slice instance (SI) use cases independently and efficiently by running all network implementations in parallel, simultaneously on a common physical network infrastructure. We devise a common shared architecture for the NH-$\mu$O small cell base stations and dynamic spectrum assignment control unit, and their required functionalities supporting coexistence of different SIs as well as multiple MNOs in shared spectrum access communications. We devise both inter-SI and intra-SI dynamic spectrum allocation policies considering time-varying requirements of different SIs. The policies are capable of taking care of application level priority, -i.e., mixture of guaranteed quality of service and best-effort service users served by each SI while ensuring a healthy competition. Our proposed framework serves two-fold advantages, such as it gives the venue owner its own managed wireless networks tailored to its very specific requirements, and it also brings out cost savings and coverage extension for MNOs and efficiency of resources that arise from sharing wireless networks, and delivering the network capacity into high density venues.
\end{abstract}

\begin{keywords}
Small cell networks, Micro operator, Shared spectrum access, 5G.
\end{keywords}
\IEEEpeerreviewmaketitle
\section{Introduction}

The use-cases for future networks are extremely diverse, for example, (i) mobile broadband experience everywhere with wide scale availability, higher speeds and more videos, (ii) ultra-reliable low latency communications, (iii) massive machine type communications, which are the official ITU-R terminology on the 5G generic services\cite{Andrews}. Furthermore, business perspective and/or context is continuously evolving. Spectrum proprietorship, network framework and content as well as customers are subject to major revolutions in near future due to advent of dense small cell networking architecture\cite{Kibria}. Industrial, enterprise and venue automation increasingly require excellent wireless connectivity.
Smart hospital, enterprises, manufacturing companies cannot not just purely rely on unlicensed wireless for their future, mission-critical connected systems. These venue owners are putting a lot of efforts on creating managed wireless networks spanning the whole venue, effectively having a robust and/or reliable networks to serve their business specific services. Moreover, the services involved may be so specific or limited to a certain location that a mobile network operator (MNO) finds it difficult or even impossible to offer them.  Traditional mobile operators cannot provide all the connectivity requirements or they are poorly-suited to providing it, and industry will not opt to pay for all wireless connectivity ``as a service", especially inside venues, buildings or across their industry/campus facilities.

The concept of micro operator ($\mu$O) has recently been proposed for local service delivery in 5G to build indoor small cell communication infrastructure and offer context related services and content in specific locations with locally issued spectrum licenses \cite{Ahokangas}. The micro operator business model is emerging from the growing number of property owners using local networks to provide specific, localized services to public buildings, shopping malls or industrial sites, and hospitals. 
A $\mu$O has own infrastructure but not necessarily own customer base. It can also support any MNO customers for efficient local service delivery. New 5G network functions become a platform for $\mu$O to take full advantage of the localized shared spectrum and telecommunication cloud resources with rapid content processing, delivering with ultra-responsive experience. Under the $\mu$O network model, the venue owners, enterprises are themselves becoming services companies, with various in-field business models related to their products and services. Note that the $\mu$O architecture consideration is dependent on several regulatory issues, such as, the availability of spectrum for local licensing for vertical specific service delivery, rules for data acquisition and utilization, rights to build indoor networks.

Since the advent of Internet, technological growth rate has been exponential in a countless number of fields, and the future is about contextual services. The venues such as shopping centers, stadiums, hospitals have their own user base as well as customers, and the success of a venue may depend critically on understanding the customers in that venue and mining their meta-data. Depending on the business types and/or purpose of the venues, there may be more varying local context related applications where the system capacity may not be the critical point, but they rather have stringent latency and/or reliability and/or massive connections requirements. In the following, we discuss two local context related use-cases and/or applications/services. The venue owners are putting a lot of efforts on making a more immersive experience from the moment the users/customers enter the venue till the moment they leave. Localized/local context related services/applications can improve venue's own regular business operations as well as can generate revenue through attracting more events and/or customers to the venues.

{\textit{Service Provision in Shopping Mall Scenario.}}
Taking into account that because the user is currently present in big shopping mall environments, relevant information and/or services can proactively be provided to the user, preferably depending on his/her preferences. When the users are detected within a certain smart shopping space, the services can be automatically discovered and presented in the mobile devices. Upon entering the new and/or unfamiliar context areas, the customers are presented contextual services such as store/first aid/emergency exit direction service, venue navigation, goods directory service, discount services, event informations through featuring specialized app. The shop can also provide excellent in-store users' experience through its own engine. Therefore, local context related services enable the venue owners with increased revenue options for creating more engaging mobile applications to enhance customer experiences and improve business revenue. 

{\textit{Service Provision in Hospital Scenario.}}
More recently, medical sensors have incorporated wireless connections, both short range to communicate wirelessly to nearby controllers/devices, and long range, such cellular communications, to communicate directly with cloud computing services. 
Some of the new medical services/applications will demand an end-to-end latency of a few milliseconds, while fields such as wireless automation and control of medical instruments/equipments may in addition require ultra reliabilities, for example, critical control of care-robots, portable bedside sensors, remote devices, surgical systems in operation theatres, etc. 
The $\mu$O network framework enables the hospital to solve these critical-wireless issues tailored to their localized services and requirements.
The hospital can also maintain a ``guest network" where patients and visitors can access the hospital network wirelessly on their personal devices for local context related services as well as for general purpose use. Guest networks are provided as a courtesy to patients and their families. Hospital guest networks can be treated differently than the main wireless network used for hospital/clinical operations, and will have different priority when the bandwidth is shared.

{\it{Shared Spectrum Access Communications}}. 
The spectrum policy should reflect future smart venues' needs and requirements. The $\mu$O needs to have the right to acquire (locally) self-controlled spectrum for managed networks for localized services. The $\mu$O architecture consideration is strongly coupled to the availability of spectrum for local licensing for vertical specific service delivery. The spectrum regulator needs to put more focus on shared spectrum access, thus allowing other stakeholders such as $\mu$O to own and operate wireless networks.
Shared spectrum access is becoming a very important component for dynamic spectrum assignment for next generation communication systems in order to achieve highly efficient spectral utilization \cite{Lin, Luo}. Due to the inadequacy of the spectrum resources and to support the imminent massive wireless demand eruption, it is important to make new spectrum available for wireless services, and make full use of the existing radio resources \cite{Staple}. Spectrum sharing presents an intermediary means to conventional exclusive licensing and license-exempt approaches, and can be put into action with the existing network infrastructure with the support of new technologies. Dynamic spectrum sharing has been found to be a very competent way to enhance the spectrum utilization efficiency and alleviating spectrum scarcity compared to the conventional exclusive licensing schemes.

An important concern of the MNOs is to minimize the capital and operational expenditures while they extend their capacity and coverage \cite{Knoll}. As growing amount of mobile traffic originates from local indoor usage, serving the high traffic hot spots with traditional cellular networks becomes insufficient as building penetration losses limit the indoor connectivity. Heterogeneous networks (HetNets) are an effective approach for the MNOs to provide the indoor users high network capacity. Furthermore, in-building/venue cellular coverage and capacity is an increasingly important component of both enterprise and residential consumers. MNOs desire to achieve improved and extended coverage, and satisfy ever-increasing capacity requirement in densely populated venues, enterprise buildings, factories, campuses, etc. Although many MNOs wish to deploy their own solutions and infrastructures in a venue, the cost of providing the additional coverage may not substantiate the benefit or revenue for improved services. As a result, the MNOs are more acceptive and considerate to business models that would endorse them to have extended coverage and improved service while keeping the capital and operational cost in check. Moreover, many venues are simply unable to accommodate separate deployments of each MNO hardware due to physical space limitations and/or appearance. It must be noted that if only one MNO is deployed in a venue, majority of the users from all other MNOs are under-supported. Therefore, more efficient business models and shared infrastructure architecture are essential to MNOs plans of providing high-quality services at any time and everywhere. One example of business models currently under consideration by MNOs is to rent local infra-structure such as base-stations and access points deployed by third-parties in private locations lacking proper MNO coverage or where increased offloading capabilities are deemed necessary.

\subsection{Contributions}

In this work, we conceive an advanced wireless networks deployment framework that facilitates $\mu$Os to become service companies. The envisioned 5G technology will enhance the venues, manufacturing and related industries in numerous ways. In this proposed framework, the $\mu$O, along with performing its localized applications, it also takes the role of a neutral host (NH)\footnote{By NH we mean a single deployment by the $\mu$O as a service provider, for example, base stations, serving users of multiple MNOs without being biased towards any of the MNOs, more likely with a single/similar radio access network with some additional functionalities to comply with the adopted sharing framework.} to serve the users from all MNOs in its vicinity for providing mobile broadband experience with a single deployment under shared spectrum access framework. Each use-case will need a distinctive configuration of requirements and parameters in the network. Thanks to radio access network slicing\footnote{The network slicing concept has originally been proposed for the 5G core network. Within 3rd generation partnership project (3GPP), partitioning of the core network is considered to be the primary target of network slicing, however, it has not excluded the notion of partitioning of resources for different network slices \cite{3GPP}.} that supports and optimizes both the use-cases efficiently \cite{Samdanis, Jiang}. Radio access network slicing is performed by breaking down the single physical infrastructure into multiple instances, which are isolated from each other and serve the use-cases separately. Thanks to software defined networking \cite{Xia,Chen}, which provides the ability to virtualize and slice the network. Our proposed framework serves two-fold advantages, such as it gives the venue owner its own managed networks tailored to its very specific requirements and brings out cost savings and coverage extension for MNOs and efficiency of resources that arise from sharing wireless networks, and delivering the network capacity into high density venues \cite{Qualcomm, Wireless2020}.

We propose the concept of neutral host-micro operator (NH-$\mu$O) for providing localized and/or local context related services to their own users belonging to slice instance I (SI-I) as well as for providing mobile broadband experience belonging to slice instance II (SI-II)  to the users from different MNOs with a single deployment under shared spectrum access framework. In principle, from the above discussions, SI-I corresponds to the services offered under purely $\mu$O framework and SI-II corresponds to the services offered by $\mu$O under NH framework. The specific contributions of this paper are:

\begin{itemize}

\item We propose an advanced shared spectrum access model under NH-$\mu$O network framework providing localized and/or local context related services as well as mobile broadband experience to the users in a venue where only a single infrastructure is mandated, for example, property owners are restrictive and prefer one infrastructure for all MNOs or the venues simply cannot accommodate separate and disjointed deployments by each MNO. NH-$\mu$O provides the $\mu$O stakeholder with a unique solution tailored to its specific needs that MNOs cannot offer\footnote{Note: both practical Wi-Fi and long term evolution-unlicensed (LTE-U) speed and distance are severely limited. Many critical security flaws are associated with Wi-Fi. Although an MNO can boost data speed over short distance with LTE-U without the need of users to log in to a different Wi-Fi network, it is not interference-free as it must accept interference due to the fact that it cannot have a vested or recognizable right to the continued usage of any unlicensed frequency. }, where the $\mu$O can address venue-specific customer needs and optimize dedicated services for end users. 

\item We devise a common shared architecture for the NH-$\mu$O small cell base station (NH-$\mu$O-SBS) and dynamic shared spectrum assignment control unit (NH-$\mu$O-SSCU), and their required functionalities that sustain coexistence of different SIs as well as multiple MNOs in shared spectrum access communications. Consequently, the venue owners under $\mu$O framework and MNOs under NH framework can have different architectural flavors for each use-case or SI or service group. NH-$\mu$O runs all network implementations in parallel, simultaneously on a common physical network infrastructure.

\item We conceive radio access network slicing concept to support and optimize both the SI use-cases independently and efficiently, and design algorithms for dynamic resource sharing across SIs as well as MNOs involving inter-SI and multi-MNO sharing policies. Our main contribution has been designing practical solutions that scale to large networks and can track network load dynamics.

\end{itemize}

\subsection{Organization}
This paper is organized as follows. Our system model based on NH-$\mu$O is described in Section II. This paper focuses on providing services to the users of multiple MNOs through a single deployment within a shared spectrum access framework, the architectures of the NH-$\mu$O-SBS and the shared spectrum access control unit along with the required functionalities for such operation are provided in Section III. We propose several spectrum coordination policies in Section IV, which can be adopted by the shared spectrum access control unit to facilitate spectrum allocation to different SIs, and and further discuss the capabilities of the spectrum coordination policies in terms of prevention of unethical exploitation of the shared spectrum access in Section V. Numerical evaluation results are given in Section VI. Concluding remarks are given in Section VII.

\section{System Model}
\label{SM}

We consider an NH-$\mu$O operated single deployment consisting of NH-$\mu$O-SBSs in a venue where separate deployments are not mandated and/or the venues that are restrictive and does not allow disjointed deployments by all the MNOs. Working under $\mu$O's NH flavor may be the only approach for the MNOs to have their footprints and in venue or enterprise customers. 
 The NH-$\mu$O exploits network slicing as a mean for reducing the capital and operations expenditures. The NH-$\mu$O network can be flexibly built in a way that coverage, capacity and speed can be allocated to different SIs to meet the specific demands. The NH-$\mu$O-SBS can serve two distinct user groups separately with different SIs. The sets of service classes (SCs) offered under the SIs are more likely to be dissimilar. The SIs are isolated and they work independently from each other.

\begin{figure}
  \centering
   \includegraphics[scale=0.08]{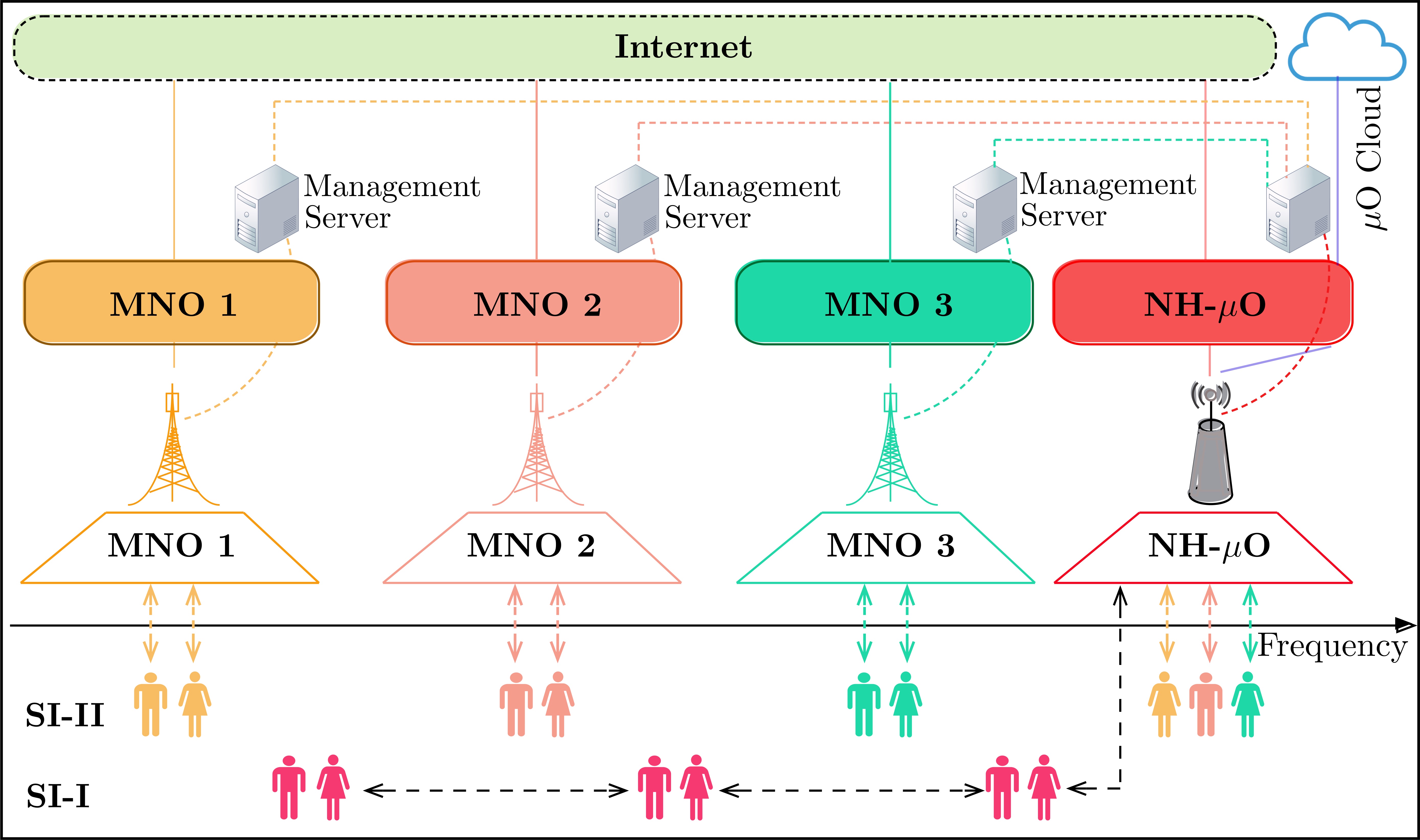}
   \caption{NH-$\mu$O based shared spectrum access communications scenario. NH-$\mu$O deployment reflects the intention of the venue owner. Users from all cooperating MNOs can attach to NH-$\mu$O from cellular services. The NH-$\mu$O can have its own user base who are provided with localized services, which are very specific to the NH-$\mu$O's business operation that the MNOs find difficult to offer.}
   \label{figb}
\end{figure}

The NH-$\mu$O deployment scenario under shared spectrum access is given in Fig.~\ref{figb}. Different colors are used to distinguish spectrums and users for different SIs and MNOs. Spectrum licenses obtained by the MNOs can be be used in an exclusive manner outside of the venue as in conventional cellular model. However, the license obtained by NH-$\mu$O is available for shared spectrum access, and acts as the basis for our spectrum sharing model. Therefore, the NH-$\mu$O supplies a common shared infrastructure that provides localized and/or local context related services to venue owners own user base as well as the users of all MNOs in a venue, with economics that favor both the MNOs and venue owner. A well-defined spectrum sharing framework manages spectrum allocation among different SIs considering time-varying requirements of different SIs. A  similar sharing framework agreed-upon by all the MNOs and NH-$\mu$O defines how the MNOs access the spectrum licensed to NH-$\mu$O in case of intra-SI spectrum allocation.

We consider a shared spectrum access model where $N_{\text{MNO}}$ MNOs along with the NH-$\mu$O participate in the spectrum auctioning process, for example, as conceived for CBRS band. The entities can be a mixture of the types of conventional cellular MNOs, mobile virtual network operators (MVNOs) and tower company. In this work, we regard the NH-$\mu$O as a third party company that has business deals with the venue owner mandating a single deployment. A sharing framework is formed that is agreed-upon by all the MNOs and NH-$\mu$O. A shared access system-spectrum coordinator (SAS-SC) manages both inter-SI and intra-SI spectrum allocation. The SIs favor each other in terms of allowing their excessive spectrums to be exploited by their counterparts when they are lightly loaded.  

\section{Proposed Architecture for NH-$\mu$O Infrastructure and the required functionalities}

Efficient architectures for NH-$\mu$O-SBS and the $\mu$O-SSCU module are very important for dynamic and adept operation of the overall shared spectrum access under NH-$\mu$O framework with only a single deployment for both local context related services as well as mobile broadband experience. 
In our considered model, we partition the overall infrastructure architecture into three modules, namely, (i) NH-$\mu$O-SBS module, (ii) NH-$\mu$O-SSCU module and (iii) NH-$\mu$O management server. We discuss the building blocks of these three modules as shown in Fig.~\ref{fig111} and their corresponding functionalities as follows.

{\textit{\textbf{Building Blocks of NH-$\mu$O-SBS and Their Functions:}}}
Apart form baseband radio frequency components and connection admission control, the NH-$\mu$O-SBS mainly has two other important components, namely, (i) Class identifier (CI) and (ii) Local resource manager (LRM) as shown in Fig.~\ref{fig111}. 
\begin{itemize}
\item CI: The CI module has two independent sub-modules, namely, CI-SI-I sub-module and CI-SI-II sub-module. The CI-SI-I sub-module associated with SI-I runs a class identification function to determine the service classes of the users served by or associated with this particular NH-$\mu$O-SBS, and reports the users's classes directly to CSM-SI-I (CSM: central spectrum manager). On the other hand, the CI-SI-I has sub-CI-SI-IIs as many as the number of MNOs under NH framework. A sub-CI-SI-II associated with an MNO runs a class identification function to determine the SCs of the users served by or associated with this particular NH-$\mu$O-SBS. The SCs identified by the sub-CI-SI-II of an NH-$\mu$O-SBS correspond to types of applications the users are running. The CI-SI-II sub-module reports the users' SCs to its corresponding local spectrum manager (LSM).

\item LRM: LRM in NH-$\mu$O-SBS is responsible for performing resource allocation, -i.e., power and resource block allocation to the data plane users associated with its NH-$\mu$O-SBS. Like CI, it also has two sub-modules, namely, LRM-SI-I and LRM-SI-II, which are working independently. The LRM-SI-I receives the assignment directly from CSM-SI-I while LRM-SI-II sub-module receives the assignment from LSMs in NH-$\mu$O-SSCU.

\end{itemize}

{\textit{\textbf{Building Blocks of NH-$\mu$O-SSCU and Their Functions:}}}
The NH-$\mu$O-SSCU unit is basically responsible for gathering informations from all CIs and LRMs. It has two important components, namely, (i) LSM and (ii) CSM.
\begin{itemize}
\item LSM: The NH-$\mu$O-SSCU has LSMs as many as the number of MNOs. One LSM is associated with one particular MNO. Each LSM manages the table of SCs offered by the MNO. Each LSM after receiving SC reports from CI-SI-II sub-modules calculates the total load of the associated MNO, and maps it to the total frequency band required for the corresponding MNO. All the LSMs report their spectrum demands to spectrum access system-spectrum manager (SAS-SC) via the central CSM-SI-II. As soon as the LSMs are notified by the CSM-SI-II about their shares in the shared spectrum access, each LSM allocates the frequency resources to the LRM-SI-II sub-modules in accordance to their demands. Note that LSM blocks are not required for SI-I.

\begin{figure}
  \centering
   \includegraphics[scale=.08]{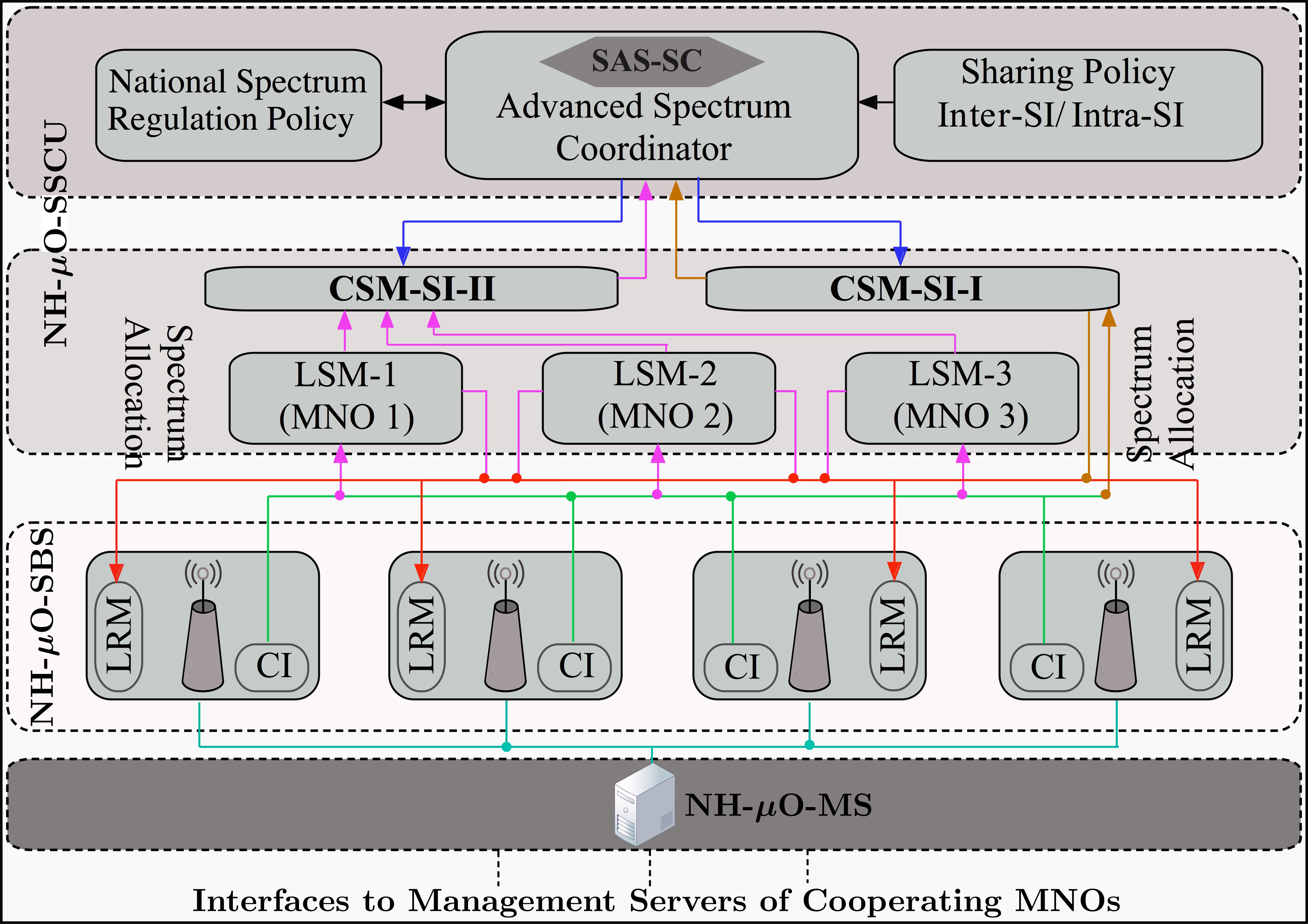}
   \caption{Building/functional blocks for common shared infrastructure of NH-$\mu$O-SBS/NH-$\mu$O-SSCU for SI-I and SI-II services. For SI-1 services, the LSM blocks are not required. The CIs can directly communicate with the CSM-SI-I which performs similar job as one of the LSMs and CSM-SI-II. }
   \label{fig111}
\end{figure}

\item CSM: CSM works as an interface between the LSMs and SAS-SC. There are two instances of CSM, namely, (i) CSM-SI-I and (ii) CSM-SI-II. The CSM-SI-II collects the spectrum demands from all the LSMs and put them forward to the SAS-SC. In case of SI-I, the corresponding CSM-SI-I collects all the informations directly from the CI-SI-I sub-modules and performs similar jobs to one of the LSMs. After SAS-SC decides the spectrum distribution, CSM-SI-II collects and reports to the LSMs, whereas CSM-SI-I can directly communicate with LRM-SI-II sub-modules.

\item SAS-SC: The SAS-SC is an advanced frequency coordinator module necessary to assign rights and maximize efficiency in the in the shared spectrum. It incorporates  environmental sensing capability (ESC), -i.e., an integrated environmental sensing capability with RF sensor networks, which is used to protect the incumbent systems via maintaining and updating the database. The SAS-SC is mainly responsible for calculating frequency resources for SIs on the basis of the frequency coordination policies discussed in Section.~\ref{SCP1}.

\end{itemize}

{\textit{\textbf{Functions of NH-$\mu$O-MS:}}} NH-$\mu$O-MS manages users inside the venue, and handles cooperation with management servers of cooperating MNOs. User authentication and mobility management scheme for handover are processed through exchanging informations between the NH-$\mu$O-MS and cooperating MNOs' management entities. NH-$\mu$O-MS most likely to be a simplified core network, and does not contain all the functionalities as in the MNOs' management servers.

\section{Proposed Shared Spectrum Access Policies}
\label{SCP}
One of the radio access network-specific requirements needed to fulfill the network slicing vision is the utilization of the radio access network resources needs to be maximized among the SIs\cite{Silva}.
Due to the time-varying requirements and workloads of the different slices, resource management is particularly important aspect in a sliced network architecture. since the slice controllers CSM-SI-I and CSM-SI-II are working independently from each other, they always try to get the required resource allocations to reach their targets. Resources are allocated to slices to guarantee their target performances if possible, or provide service differentiation if the total requests from all the slices exceed the capacities of the shared resources.

The SAS-SC module enables the dynamic resource allocation between the different SIs running on top of the physical resources. The SAS-SC collects the resource allocation requests from the different SIs using the resources it controls and determines the allocations based on the available capacities of the resources. 

{\bf{{\textit{SIs' Spectrum Demand Estimations:}}}} We consider that SI-I services, -i.e., business specific wireless services and/or the local context related services may be categorized under a number of SCs if there are more than one services, and the sets of SCs for different venues may differ depending on the business types and requirements. The SI-II offers a combination of real-time and best-effort services under mobile broadband use-case, and the offered services can be classified under different SCs. However, for traffic load estimation/prediction in the following, we do not specify any SC for SI-I services. Consequently, we consider that the SIs have dissimilar traffic patterns.

\begin{itemize}

\item {\bf{SI-I Spectrum Demand Estimation}}: For SI-I, we consider a scenario where for the $\mu$O's own regular business operation, which can be mission critical, ultra-reliable, the required traffic is fixed\footnote{Therefore, provides a dedicated connection insuring the availability of required bandwidth for interference-free, ultra-reliable and uninterrupted connectivity.} to $T_{{\rm{SI-I}}}$ Mbps while for other local context related services to its visitors or its own users, the traffic follows a uniform distribution defined as $\mathcal{U}[T_{\text{min}},\hspace{1mm}T_{\text{max}}]$, where $T_{\text{min}}$ and $T_{\text{max}}$ are lower and upper limits of the traffic distribution, respectively. 
The CSM-SI-I employs a simple mapping function given by 
\begin{equation}
\text{Map}_{f-{\rm{SI-I}}}:\hspace{1mm}T_{{\rm{SI-I}}}+R_{{\rm{SI-I}}}\xrightarrow{/\eta_{\rm{bps/Hz}}}\Delta_{{\rm{SI-I}}},
\label{eq3}
\end{equation}
to estimate its spectrum demand from its calculated load, where, $\eta_{\rm{bps/Hz}}$ is the spectral efficiency in bps/Hz.

\item {\bf{SI-II Spectrum Demand Estimation}}:  Without loss of generality, we consider that all the MNOs operating under SI-II in shared spectrum access offer services belonging to the same set of SCs. Let $u_q$ be the number of users served by MNO $q$. The set of SCs offered by any MNO can be denoted as $\mathfrak{S}_{\rm{SI-II}}\triangleq$ \{$\mathbb{VL}$, $\mathbb{L}$, $\mathbb{M}$, $\mathbb{MH}$, $\mathbb{H}$, $\mathbb{S}$\}, where $\mathbb{VL}$, $\mathbb{L}$, $\mathbb{M}$, $\mathbb{MH}$, $\mathbb{H}$ and $\mathbb{S}$ stand for SCs corresponding to very-low, low, medium, medium-high, high and super high data-rate applications, respectively. Each SC is associated with a given data-rate in Mbps. Each connection, -i.e., the connection between any NH-$\mu$O-SBS and the user is associated with a priority/weight that reflects its SC, which is a function of maximum outage probability, maximum allowable delay, virtual waiting time, target error rate, etc. The priority/weight parameters can be estimated by the extended EXP method \cite{Shakkottai, Wang}. Let, $\omega_{s}$ be the priority parameter of the users belonging to SC $s$. Considering both the user and/or device activity and service usage patterns, the traffic load calculation performed by LSM $q$ is given as

\begin{equation}
R_q=\sum_{s\in\mathfrak{S}_{\rm{SI-II}}}\omega_{s}R_{q,s} \text{ Mbps}
\end{equation}
where, the load for a specific SC $s$ on MNO $q$, $R_{q,s}$ is given by
\begin{equation}
R_{q,s}=\left(u_q\times \alpha_f^q\right)\times \beta_f^{qs}\times \delta_s^{{\rm{SI-II}}} \text{ Mbps}.
\end{equation}
Here, $\alpha_f^q$ is the activity factor ($\%$ active users). The parameter $\beta_f^{qs}$ is the service usage patterns which defines the number of users using a particular SC $s$ under MNO $q$, and the parameter $\delta_s^{{\rm{SI-II}}}$ is the application data-rate that corresponds to the SC $s\in\mathfrak{S}_{\rm{SI-II}}$. The CSM-SI-II aggregates the demands from all the LSMs as $R_{\text{SI-II}}=\sum_{q=1}^{N_{\text{MNO}}}R_q$ and employs a simple mapping function given by 
\begin{equation}
\text{Map}_{f-\text{SI-II}}:\hspace{1mm}R_{\text{SI-II}}\xrightarrow{/\eta_{\rm{bps/Hz}}}\Delta_{\text{SI-II}},
\end{equation}
to estimate its spectrum demand from its calculated load.

\end{itemize}

\subsection{Inter-SI Spectrum Allocation}
\label{SCP1}
SAS-SC manages the distribution of network resources among slices, and adopts a fair, efficient and transparent process for awarding spectrum licences for different SIs services.
In this section, we discuss our proposed spectrum coordination policies employed by the SAS-SC under NH-$\mu$O framework to dynamically assign the shareable spectrum among SIs. We propose a pool of three spectrum coordination policies namely (i) Fixed reserved (FR) mode (ii) Primary share reserved (PR) mode and (iii) Complete sharing (CS) mode. The best policy can be adopted by SAS-SC conditioned on mutual agreement by the MNOs and the $\mu$O. The proposed policies are discussed in details in the following subsections. Let us define $\mathcal{S}\triangleq\{\text{SI-I, SI-II}\}$. The sharing framework is formed based on the dedicated band of $\mathfrak{B}_{\rm{T}}$ MHz to the NH-$\mu$O. Let $\mathcal{B}_s=\mathfrak{R}_s\mathfrak{B}_{\rm{T}}, s\in\mathcal{S}$ ($0<\mathfrak{R}_s<1$) be the amount of spectrum the SI $s$ obtains after SAS-SC runs our proposed spectrum coordination policy, where $\mathfrak{R}_s\times 100\%$ reflects SI $s$'s ultimate share in percentage.

\subsubsection{FR Mode-Guaranteed Allocation}
In this mode of spectrum allocation, both the slices have reserved spectrum for their operations under NH-$\mu$O. Therefore, irrespective of the loads and/or demands of the slices, they have exclusive access to their respective fully reserved spectrum. As a result, dynamic spectrum sharing does not happen in FR-mode, -i.e., FR mode allows purely a static spectrum allocation, where the parameter $\mathfrak{R}_s$ is given by

\begin{equation}
\mathfrak{R}_s=\psi_s, \hspace{1mm}\forall s, s\in\mathcal{S},
\end{equation}
with $\psi_{\text{SI-I}}+\psi_{\text{SI-II}}=1$. Consequently, $\mathcal{B}_{\text{SI-I}}\cup\mathcal{B}_{\text{SI-I}}=\mathfrak{B}_{\rm{T}}$. In other words, we have $\mathcal{B}_{\text{SI-I}}:\mathcal{B}_{\text{SI-II}}=\psi_{\text{SI-I}}:\psi_{\text{SI-II}}$ irrespective of the demands from SIs. The values of $\psi_{\text{SI-I}}$ and $\psi_{\text{SI-II}}$ are decided through mutual consent in an agreement form. Therefore, in FR mode, there are pre-determined rigid boundaries among the frequency bands allocated to different SIs. It is the most inefficient spectrum sharing approach among all three modes. The SIs do not favor each other in terms of allowing other SIs to utilize its excessive spectrum when it is lightly loaded, -i.e., when the spectrum is underutilized. This is the same as conventional exclusively licensed spectrum allocation in cellular systems.

\subsubsection{PR Mode-Minimum Guaranteed Allocation}
The proposed spectrum coordination policy based on PR mode facilitates the SIs to have a minimum guaranteed allocation if they require. In this mode, the SIs have their principal shares $\{\xi_{\text{SI-I}},\xi_{\text{SI-II}}\}$ MHz agreed-upon depending on number of users, reliability, latency requirements and supported sets of service classes. Principal share defines the percentage of shareable spectrum an SI is assured when all the SIs are overloaded or underloaded, -i.e., defines the minimum guaranteed allocation if required. Summation of primary shares may be less than or equal to the total shareable spectrum $\mathfrak{B}_{\rm{T}}$. If the summation is less than $\mathfrak{B}_{\rm{T}}$, then we have $\xi_{\text{SI-I}}+\xi_{\text{SI-II}}+\Theta=\mathfrak{B}_{\rm{T}}$, where $\Theta$ is the amount of reserved spectrum. Therefore, if $\Theta=0$, we have $\xi_{\text{SI-I}}+\xi_{\text{SI-II}}=\mathfrak{B}_{\rm{T}}$. The CSM-SI-I and CSM-SI-II sub-modules associate their spectrum demands with a priority parameter $\{\rho_{\text{SI-I}},\rho_{\text{SI-II}}\}$ which can take values from a predetermined set of values, or are fixed through the agreement. An SI with higher priority will have an edge over the other SI in terms of gaining more spectrum (from the excessive/superfluous spectrum if there is any) to serve its users. The PR mode based spectrum coordination mechanism provided in {\bf{Algorithm \ref{ALG1}}} works as follows.
\begin{algorithm}
\label{ALG1}
\caption{PR Mode: Inter-SI Spectrum Coordination}\label{alg:euclid}
\begin{algorithmic}[1]
\State Inputs: $\{\xi_{\text{SI-I}},\xi_{\text{SI-II}}\}$, $\Theta$, $\mathfrak{B}_{\rm{T}}$, $\{\rho_{\text{SI-I}},\rho_{\text{SI-II}}\}$;
\State Normalize: $\bar{\rho_s}=\rho_s/\left(\rho_{\text{SI-I}}+\rho_{\text{SI-II}}\right)$;
\State Normalize: $\bar{\xi_s}= \xi_s/\left(\xi_{\text{SI-I}}+\xi_{\text{SI-II}}+\Theta\right)$;
\State Define: \hspace{.8mm}  $\mathfrak{N}_{+}\triangleq\{s\hspace{1mm} |\hspace{1mm} \Delta_s-\bar{\xi_s}\mathfrak{B}_{\rm{T}}\le0\}$;
\State \hspace{12.4mm} $\mathfrak{N}_{-}\triangleq\{s\hspace{1mm}|\hspace{1mm}\Delta_s-\bar{\xi_s}\mathfrak{B}_{\rm{T}}>0\}$;
\State {\bf{C-I}}: $\Delta_s\le \bar{\xi_s} \mathfrak{B}_{\rm{T}}, \forall s,\hspace{1mm} s\in\mathcal{S} \text{ or }$ $\sum_{s\in\mathcal{S}} \left(\Delta_s-\bar{\xi_s}\mathfrak{B}_{\rm{T}}\right)\le0$

\State \hspace{0mm} Set: $\mathfrak{R}_s=\Delta_s/\mathfrak{B}_{\rm{T}}$;
\State {\bf{C-II}}:  $\Delta_s> \bar{\xi_s}\mathfrak{B}_{\rm{T}} ,  \forall s,\hspace{1mm} s\in\mathcal{S}$
\State {\bf{if}} $\Theta=0$
\State \hspace{0mm} Set: $\mathfrak{R}_s=\xi_s$;
\State {\bf{if}} $\Theta\ne0\hspace{1mm} \&\hspace{1mm} \sum_{s\in\mathcal{S}} \left(\Delta_s-\bar{\xi_s}\mathfrak{B}_{\rm{T}}\right)\le\Theta$
\State \hspace{0mm} Set: $\mathfrak{R}_s=\Delta_s/\mathfrak{B}_{\rm{T}}$;
\State {\bf{if}} $\Theta\ne0\hspace{1mm} \&\hspace{1mm} \sum_{s\in\mathcal{S}} \left(\Delta_s-\bar{\xi_s}\mathfrak{B}_{\rm{T}}\right)>\Theta$
\State \hspace{0mm} Set: $\mathfrak{R}_s=\left(\bar{\xi_s}\mathfrak{B}_{\rm{T}}+\Theta\times\bar{\rho_s}\right)/\mathfrak{B}_{\rm{T}}$; 
\State {\bf{C-III}}: $\sum_{s\in\mathcal{S}} \left(\Delta_s-\bar{\xi_s}\mathfrak{B}_{\rm{T}}\right)>0$ 
\State {\bf{if}} $\Theta=0$
\State \hspace{0mm} Set: $\mathfrak{R}_s=\Delta_s/\mathfrak{B}_{\rm{T}}, s\in \mathfrak{N}_{+}$;
\State \hspace{0mm} Set: $\mathfrak{R}_s=\left(\bar{\xi_s}\mathfrak{B}_{\rm{T}}+\sum_{j\in\mathfrak{N}_{+}} \left (\Delta_j-\bar{\xi_j}\mathfrak{B}_{\rm{T}}\right )\right)/\mathfrak{B}_{\rm{T}}, s\in \mathfrak{N}_{-};$
\State {\bf{if}} $\Theta\ne0$ $\&$ $\sum_{s\in\mathcal{S}} \left(\Delta_s-\bar{\xi_s}\mathfrak{B}_{\rm{T}}\right)\le\Theta$
\State \hspace{0mm} Set: $\mathfrak{R}_s=\Delta_s/\mathfrak{B}_{\rm{T}}, s\in \mathcal{S}$;
\State {\bf{if}} $\Theta\ne0$ $\&$ $\sum_{s\in\mathcal{S}} \left(\Delta_s-\bar{\xi_s}\mathfrak{B}_{\rm{T}}\right)>\Theta$
\State \hspace{0mm} Set: $\mathfrak{R}_s=\Delta_s/\mathfrak{B}_{\rm{T}}, s\in \mathfrak{N}_{+}$;
\State \hspace{0mm} Set: $\mathfrak{R}_s=\left(\bar{\xi_s}\mathfrak{B}_{\rm{T}}+\sum_{j\in\mathfrak{N}_{+}} \left (\Delta_j-\bar{\xi_j}\mathfrak{B}_{\rm{T}}\right )+\Theta\right)/\mathfrak{B}_{\rm{T}}, s\in \mathfrak{N}_{-};$
\end{algorithmic}
\end{algorithm}

 According to the proposed PR mode policy, at first, traffic loads calculated by the CSM-SI-I and CSM-SI-II are mapped to the corresponding spectrum demands through the mapping function in \eqref{eq3}. We break the PR policy into three distinct cases, denoted by {\textbf{C-I}}, {\textbf{C-II}} and {\textbf{C-III}}, respectively, depending on the relationship between SIs' demands and principal shares $\{\xi_{\text{SI-I}},\xi_{\text{SI-II}}\}$, and $\Theta$. 
 
 {\bf{C-I}} resembles the scenario where the demands from both the SIs are less then or equal to their principal shares and/or where the overall system is lightly loaded. If both the SIs are {\it{lightly loaded}}, -i.e., $\Delta_s\le \bar{\xi_s} \mathfrak{B}_{\rm{T}}, \hspace{1mm}s\in\mathcal{S}$ or if the demands from the SIs are such that the {\it{collective system is lightly loaded}}\footnote{One SI demands less than its principal shares, but the other SI demands higher than its principal shares. The combined demand is less than the combined principal share.}, -i.e., $\sum_{s\in\mathcal{S}} \left(\Delta_s-\bar{\xi_s}\mathfrak{B}_{\rm{T}}\right)<0 $ or if the demands from the SIs are such that the system has {\it{jointly balanced load}}, -i.e., $\sum_{s\in\mathcal{S}} \left(\Delta_s-\bar{\xi_s}\mathfrak{B}_{\rm{T}}\right)=0$, the SAS-SC assigns the SIs spectrum as much as they demand. As a results,  $\mathfrak{R}_s=\Delta_s/\mathfrak{B}_{\rm{T}}$. Note that under  {\bf{C-I}}, the priority parameters do not have any influence on the spectrum allocation.

 {\bf{C-II}} resembles a shared spectrum access scenario where the demands from both the SIs are outside their limits, -i.e., $\Delta_s> \bar{\xi_s}\mathfrak{B}_{\rm{T}} ,  \forall s,\hspace{1mm} s\in\mathcal{S}$, therefore, both the SIs are {\it{heavily loaded}}. In such a scenario, there appears three different allocation scenarios depending on the value of $\Theta$. For example, if $\Theta=0$, -i.e., there is no reserved spectrum, the SIs just receive their principal shares. Whereas, if $\Theta\ne0$ and $\Theta$ is less than or equal to the combined excessive demand, then the SIs have access to the reserved band and receives as much as they demand. Finally, if $\Theta\ne0$ and $\Theta$ is less then combined excessive demand, the priority parameters play their roles in deciding the ultimate spectrum sharing between the SIs. In this situation, the reserved spectrum is proportionally distributed to the SIs, the parameter $\mathfrak{R}_s$ is given  by
 \begin{equation}
 \mathfrak{R}_s=\left(\bar{\xi_s}\mathfrak{B}_{\rm{T}}+\Theta\times\bar{\rho_s}\right)/\mathfrak{B}_{\rm{T}}.
 \end{equation}
 
 Finally,  {\bf{C-III}} depicts a scenario, where the system has {\it{jointly unbalanced load}} while the overall system is {\it{heavily loaded}}, -i.e., $\sum_{s\in\mathcal{S}} \left(\Delta_s-\bar{\xi_s}\mathfrak{B}_{\rm{T}}\right)>0$. Similar to  {\bf{C-II}}, we encounter three different spectrum allocation scenarios depending on the value of $\Theta$. If $\Theta=0$, -i.e., when there is no reserved spectrum, the SI demanding lower than or equal to its principal share gets whatever it demands. However, the SI, which has higher spectrum demand compared to its principal share, is favored to have access to the excessive spectrum that becomes available from the lightly loaded SI. The corresponding values of $\mathfrak{R}_s$ are give by
 \begin{equation}
\label{main6}
\mathfrak{R}_s= \left\{ \begin{array}{*{35}{l}}
\Delta_s/\mathfrak{B}_{\rm{T}}, s\in \mathfrak{N}_{+}. \vspace{2mm} \\
\left(\bar{\xi_s}\mathfrak{B}_{\rm{T}}+\sum_{j\in\mathfrak{N}_{+}} \left (\Delta_j-\bar{\xi_j}\mathfrak{B}_{\rm{T}}\right ) \right)/\mathfrak{B}_{\rm{T}},\hspace{2mm}, s\in \mathfrak{N}_{-}. \\
\end{array}\right.
\end{equation}
In case of $\Theta\ne0$ and the excessive demand from {\it{heavily loaded}} SI being less than or equal to $\Theta$, -i.e., $\sum_{s\in\mathcal{S}} \left(\Delta_s-\bar{\xi_s}\mathfrak{B}_{\rm{T}}\right)\le\Theta$, the SIs are entitled to gain spectrum as much as they demand as the excessive demand from the {\it{heavily loaded}} SI comes from the reserved spectrum. Finally, under the scenarios, where $\Theta\ne0$ and $\sum_{s\in\mathcal{S}} \left(\Delta_s-\bar{\xi_s}\mathfrak{B}_{\rm{T}}\right)>\Theta$, the values of $\mathfrak{R}_s$ are given by
\begin{equation}
\label{main6}
\mathfrak{R}_s= \left\{ \begin{array}{*{35}{l}}
\Delta_s/\mathfrak{B}_{\rm{T}}, s\in \mathfrak{N}_{+}. \vspace{2mm} \\
\left(\bar{\xi_s}\mathfrak{B}_{\rm{T}}+\sum_{j\in\mathfrak{N}_{+}} \left (\Delta_j-\bar{\xi_j}\mathfrak{B}_{\rm{T}}\right )+\Theta\right)/\mathfrak{B}_{\rm{T}}, s\in \mathfrak{N}_{-}., \\
\end{array}\right.
\end{equation}
 which states that the {\it{lightly loaded}} SI receives what it demands while the {\it{heavily loaded}} SI is allowed to full access to the excessive spectrum from {\it{lightly loaded}}
 SI as well as the whole reserved spectrum.

\subsubsection{CS Mode-Flexible Allocation}

In this mode, we consider that the SIs are allocated with spectrums that are proportional to their demands, -i.e., demand and supply are directly proportional while the total supply remains fixed.
\begin{equation}
\mathfrak{R}_s=\Delta_s/\left(\Delta_{\text{SI-I}}+\Delta_{\text{SI-II}}\right).
\end{equation}
To be specific, the SAS-SC divides the spectrum resources among the SIs in a ratio equal to the demand, therefore, yields a linear-proportional fairness criterion. However, there is no guarantee that the spectrum demands of the SIs can always be fulfilled.

\begin{figure}
  \centering
   \includegraphics[scale=.08]{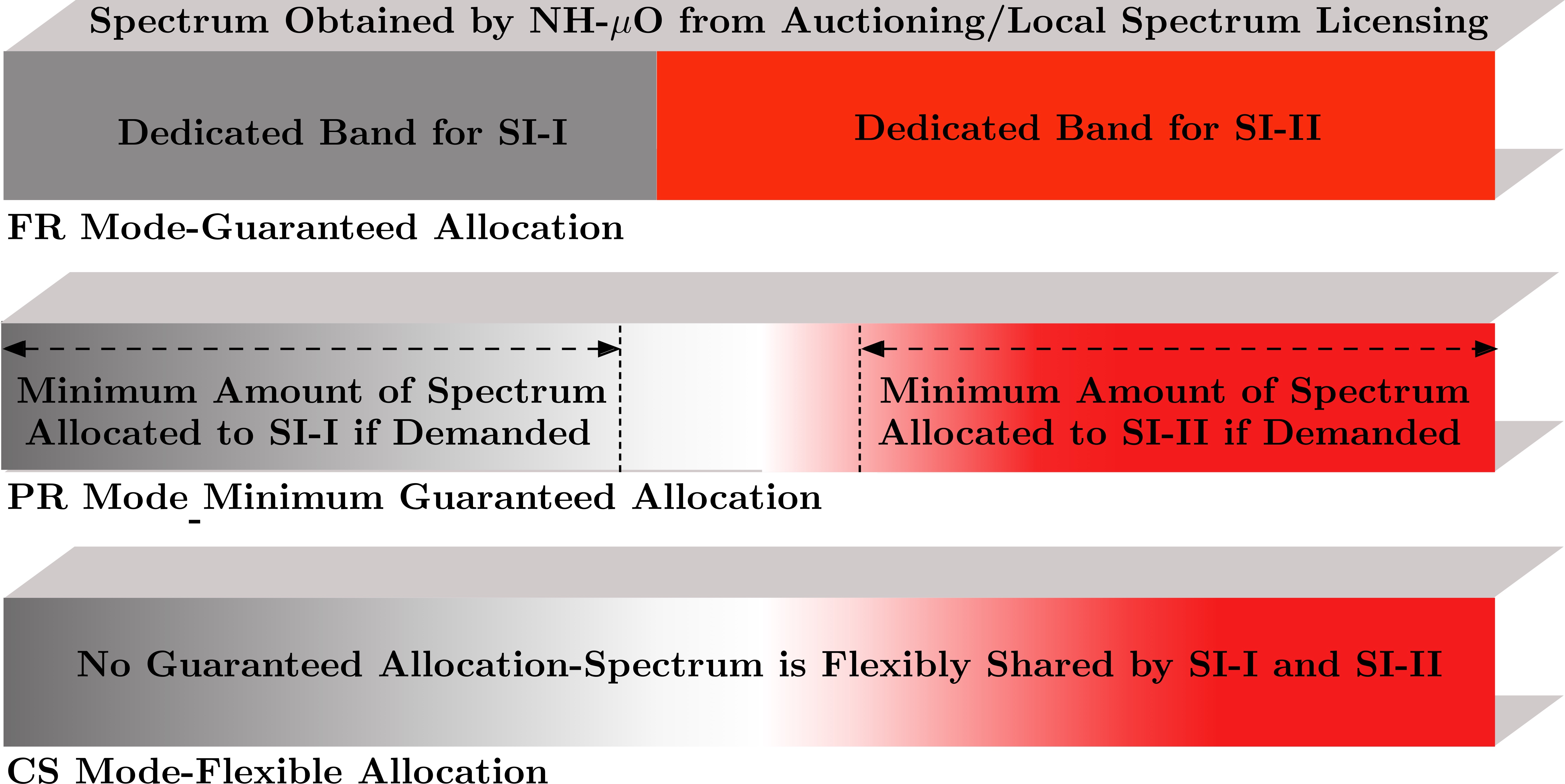}
   \caption{Schematic illustration of the spectrum coordination policies for NH-$\mu$O based shared spectrum access framework.}
   \label{fig1}
\end{figure}

\subsection{Intra-SI Spectrum Allocation (SI-II)}

For intra-SI spectrum allocation, the SAS-SC can follow the same policies discussed for inter-SI spectrum allocation. In this section, we basically discuss the intra-SI spectrum allocation for SI-II while the same approaches can be followed for SI-I. Let $\mathfrak{B}_{\text{SI-II}}$ is the amount of spectrum the SI-II is assigned with in the inter-SI spectrum allocation phase. In the following subsection we just discuss the PR mode based allocation as the other modes can readily be followed for intra-SI spectrum allocation.

\subsubsection{PR Mode-Minimum Guaranteed Allocation}
 In this mode, like the SIs, the MNOs have their principal shares $\{\xi_1,\xi_2, \cdots, \xi_N\}\%$ agreed-upon depending on sizes of subscriptions and/or market shares of the operators. Principal share\footnote{Under all the spectrum coordination policies discussed here, anti-competitive behaviour, in the form of acquisition of disproportionate/excessive spectrum, can be prohibited in multiple ways by the SAS-SC and/or regulatory authority by setting spectrum cap, which defines the limit of spectrum an MNO can possess for supporting services to its subscribers. Spectrum cap rules counter an MNO to prevent it from acquiring disproportional amount of spectrum compared to its competitors thereby not allowing wireless market domination. } defines the percentage of shareable spectrum an MNO is assured when all the MNOs are overloaded or underloaded, -i.e., defines the minimum guaranteed allocation if required. Apart from incorporating priority to the services, -i.e., to the users served by the MNOs as described in Section.~\ref{SM}, the LSMs also associate their spectrum demands with a priority parameter $\{\rho_1,\rho_2, \cdots, \rho_N\}$ which can take values from a predetermined set of values.  Any operator with higher priority will have an edge over its competitors in terms of gaining more spectrum (from the excessive/superfluous spectrum if there is any) to serve more end-users and/or to provide guaranteed QoS, and to make more profit. The operator who desires higher priority, for instance, to guarantee more spectrum to its customers,
need to reimburse through a conformed fee proportional to the degree of achieved prioritization in comparison with its competitors. In our proposed method, the user-level priority, -i.e., application-level priority is maintained/ensured by NH-$\mu$O-SBS and the upper-tier priority is ensured by the SAS-SC employing our proposed PR mode policy that establishes both short-term and long-term fairness/priority. For example, a larger carrier with massive subscriptions would more likely require larger amount of spectrum compared to a moderate or small operator with small subscriptions. Even a small operator can opt for higher priority if it wants to provide its users a better QoS. The mechanism  provided in {\bf{Algorithm \ref{ALG2}}} works as follows.
\begin{algorithm}
\label{ALG2}
\caption{PR Mode: Intra-SI Spectrum Coordination}\label{alg:euclid}
\begin{algorithmic}[1]
\State Inputs: $\{\xi_1, \cdots, \xi_{N_{\text{MNO}}}\}$, $\Theta_{\rm{SI}}$, $\mathfrak{B}_{\rm{SI}}$, $\{\rho_1, \cdots, \rho_{N_{\text{MNO}}}\}$
\State Normalize: $\bar{\xi_q}= \xi_q/(\xi_1+\xi_2+\cdots+\xi_{N_{\text{MNO}}}), \forall q,\hspace{1mm} q\in\mathcal{Q}$;
\State Normalize: $\bar{\rho_q}=\rho_q/(\rho_1+\rho_2+\cdots+\rho_{N_{\text{MNO}}}), \forall q,\hspace{1mm} q\in\mathcal{Q}$;
\State Define: \hspace{.8mm}  $\mathfrak{D}_{+}\triangleq\{q\hspace{1mm} |\hspace{1mm} \Delta_q-\bar{\xi_q}\mathfrak{B}_{\rm{SI}}\le0\}$;
\State \hspace{12.4mm} $\mathfrak{D}_{-}\triangleq\{q\hspace{1mm}|\hspace{1mm}\Delta_q-\bar{\xi_q}\mathfrak{B}_{\rm{SI}}>0\}$;
\State {\bf{C-III}}: $\sum_{q\in\mathcal{Q}} \left(\Delta_q-\bar{\xi_q}\mathfrak{B}_{\rm{SI}}\right)>0$ 
\State {\bf{if}} $\Theta=0$
\State \hspace{0mm} $\mathfrak{R}_q=\Delta_s/\mathfrak{B}_{\rm{SI}}, q\in \mathfrak{D}_{+}$;
\State \hspace{0mm} $\mathfrak{R}_q=\left(\bar{\xi_q}\mathfrak{B}_{\rm{SI}}+\frac{\rho_q}{\sum_{j\in\mathfrak{D}_{+}}\rho_q}\sum_{j\in\mathfrak{D}_{+}} \left (\Delta_j-\bar{\xi_j}\mathfrak{B}_{\rm{SI}}\right )\right)/\mathfrak{B}_{\rm{SI}}, q\in \mathfrak{D}_{-};$
\State {\bf{if}} $\Theta\ne0$ $\&$ $\sum_{q\in\mathcal{Q}} \left(\Delta_q-\bar{\xi_q}\mathfrak{B}_{\rm{SI}}\right)\le\Theta$
\State \hspace{0mm} $\mathfrak{R}_q=\Delta_s/\mathfrak{B}_{\rm{SI}}, q\in \mathcal{S}$;
\State {\bf{if}} $\Theta\ne0$ $\&$ $\sum_{q\in\mathcal{Q}} \left(\Delta_s-\bar{\xi_q}\mathfrak{B}_{\rm{SI}}\right)>\Theta$
\State \hspace{0mm} $\mathfrak{R}_q=\Delta_s/\mathfrak{B}_{\rm{SI}}, q\in \mathfrak{D}_{+}$;
\State \hspace{0mm} $\mathfrak{R}_q=\left(\bar{\xi_q}\mathfrak{B}_{\rm{SI}}+\frac{\rho_q}{\sum_{j\in\mathfrak{D}_{+}}\rho_q}\left(\sum_{j\in\mathfrak{D}_{+}} \left (\Delta_j-\bar{\xi_j}\mathfrak{B}_{\rm{SI}}\right )+\Theta\right)\right)/\mathfrak{B}_{\rm{T}}, q\in \mathfrak{D}_{-};$
\end{algorithmic}
\end{algorithm}

We again have three distinct allocation cases depending on the relationship between SIs' demands and principal shares $\{\xi_1,\xi_2, \cdots, \xi_{N_{\text{MNO}}}\}$, and $\Theta$. The spectrum allocation policies proposed under {\textbf{C-I}} and {\textbf{C-II}}  are readily applicable to PR mode based intra-SI spectrum allocation. We just discuss {\textbf{C-III}} in details in the following. 
 Under {\bf{C-III}}, where the system has jointly unbalanced load while the overall system is {\it{heavily loaded}}, we further can break {\bf{C-III}} into three different regions just depending on the value of the parameter $\Theta$. Note that these are very simple variations in the values of $\mathfrak{R}_q$ depending on the number of MNOs. If $N_{\text{MNO}}=2$, then we can just follow the procedures described in PR mode based inter-SI spectrum allocation. When we have $N_{\text{MNO}}>2$, the SAS-SC may encounter a situation where two MNOs are {\it{heavily loaded}}. In this case, the excessive spectrum that comes from {\it{lightly loaded}} MNO as well as the reserved spectrum (if $\Theta\ne0$) can be distributed among the {\it{heavily loaded}} MNOs proportionally depending on their priority parameters.

\section{Simulation Results and Performance Analysis}

For performance analysis of NH-$\mu$O with the proposed spectrum coordination policies, in the simulation, we consider an NH-$\mu$O based small cell network with two SIs, one providing local context related services and the other one providing mobile broadband experience. Let, the amount of locally licensed spectrum dedicated to NH-$\mu$O is 30 MHz which is contiguous. Without any loss of generality, we consider one-to-one mapping with BPSK modulation employed by both the SIs, -i.e., maximum throughput of 30 Mbps can be achieved. The SIs have different traffic patterns. For SI-I, we consider a scenario where for the $\mu$O's own business operation, the required traffic is fixed to 5 Mbps while for the local context related services to its visitors or its own users, the traffic follows a uniform distribution $\mathcal{U}$[1, 20] Mbps. For SI-II, we consider a conventional cellular user distribution and service types scenario, where the MNOs have varying number of users, -i.e., uniformly distributed pseudonumber integer within the range defined by [1, $u_{\text{max}}$]. The service classes/types associated with the users are also chosen randomly from the set  $\mathfrak{S}_{\rm{SI-II}}$, where the set of data-rates $\{5000, \hspace{1mm} 20000 \hspace{1mm}, 30000, \hspace{1mm}300000, \hspace{1mm} 600000, \hspace{1mm} 940000\}$ Kbps corresponds to the SCs defined in $\mathfrak{S}_{\rm{SI-II}}$. As a result, the SIs have different traffic patterns.

The load of each SI as well as MNO (under SI-II) varies over time, and as a result, the spectrum demands from different SIs (also MNOs) are temporally independent. Depending on the instantaneous demands of the SIs, the SAS-SC assigns spectrums according to the spectrum coordination policies described in Section.~\ref{SCP1}. 
The deviation parameter $\epsilon_p$, which denotes the absolute difference between the demand and supply for the SIs given by $|\Delta_{p}-\mathfrak{B}_p|, \hspace{1mm}p\in\mathcal{S}$, is used as a performance measure of the spectrum coordination policies. From the definition of the deviation parameter, the lower the value of $\epsilon_p$, the better the performance. 
We consider that the SAS-SC makes spectrum coordination decisions every $T$ time window. Every $T$ time window, the CSM-SI-I and CSM-SI-II present their resource request based on traffic estimation/prediction.

\begin{figure}
\centering
\includegraphics[scale=.08]{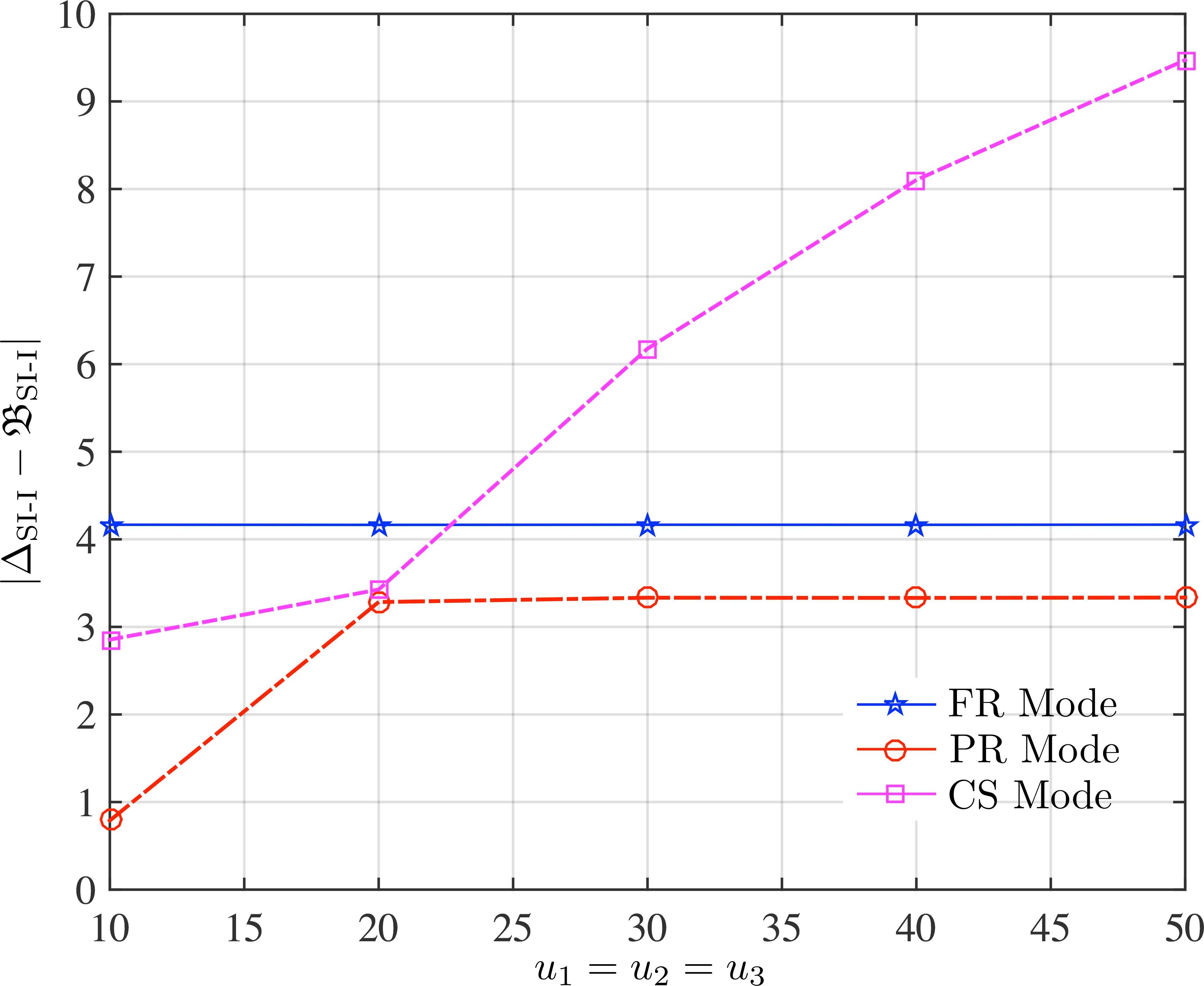}
\caption{Deviation characteristics for SI-I with different modes of spectrum coordination policies when the number of users belonging to the MNOs under SI-II varies. Here, $\xi_{\text{SI-I}}=\xi_{\text{SI-II}}=0.5$ and $\Theta=0$ MHz.}
\label{perffig1}
\end{figure}
In Fig.~\ref{perffig1}, we investigate the characteristics of the deviation measure $|\Delta_{p}-\mathfrak{B}_p|$ of SI-I for all the considered spectrum coordination policies when the number of users belonging to the MNOs under SI-II varies. The traffic of SI-I follows uniform distribution $\mathcal{U}$[1, 20] Mbps. For the PR mode, the SIs' principal shares are taken as $\bar{\xi}_{\text{SI-I}}=\bar{\xi}_{\text{SI-I}}=0.5$. It can be seen that the FR mode exhibits consistent and/or equal deviation performance even when there is a load variation on SI-II and it is obvious. This is because, no matter what the load is on SI-II, SI-I gets its principal share independent of its own load. It should also be noted that SI-I experiences the largest deviation if FR mode is employed compared to PR and CS mode when the load on SI-II is comparatively small. On the other hand, as the load on SI-II increases, the CS mode happens to provide increasing deviation to SI-I as the available spectrum is proportionally distributed between the SIs depending on the load. As the load on SI-II increases, the SAS-SC assigns more and more spectrum to SI-II, which results in higher deviation measure for SI-I. Therefore, If the number of users per MNO under SI-II increases and/or more users are using applications that require higher bandwidth, then the spectrum demand of the SI-II will also increase. However, as the bandwidth of the whole system is limited, a bottleneck occurs. The PF mode is found to be the best spectrum allocation policy in this scenario as PF mode allows mutual favoring. The unused spectrum of one SI can be used by the other SI if it requires.

\begin{figure}
\centering
\includegraphics[scale=.08]{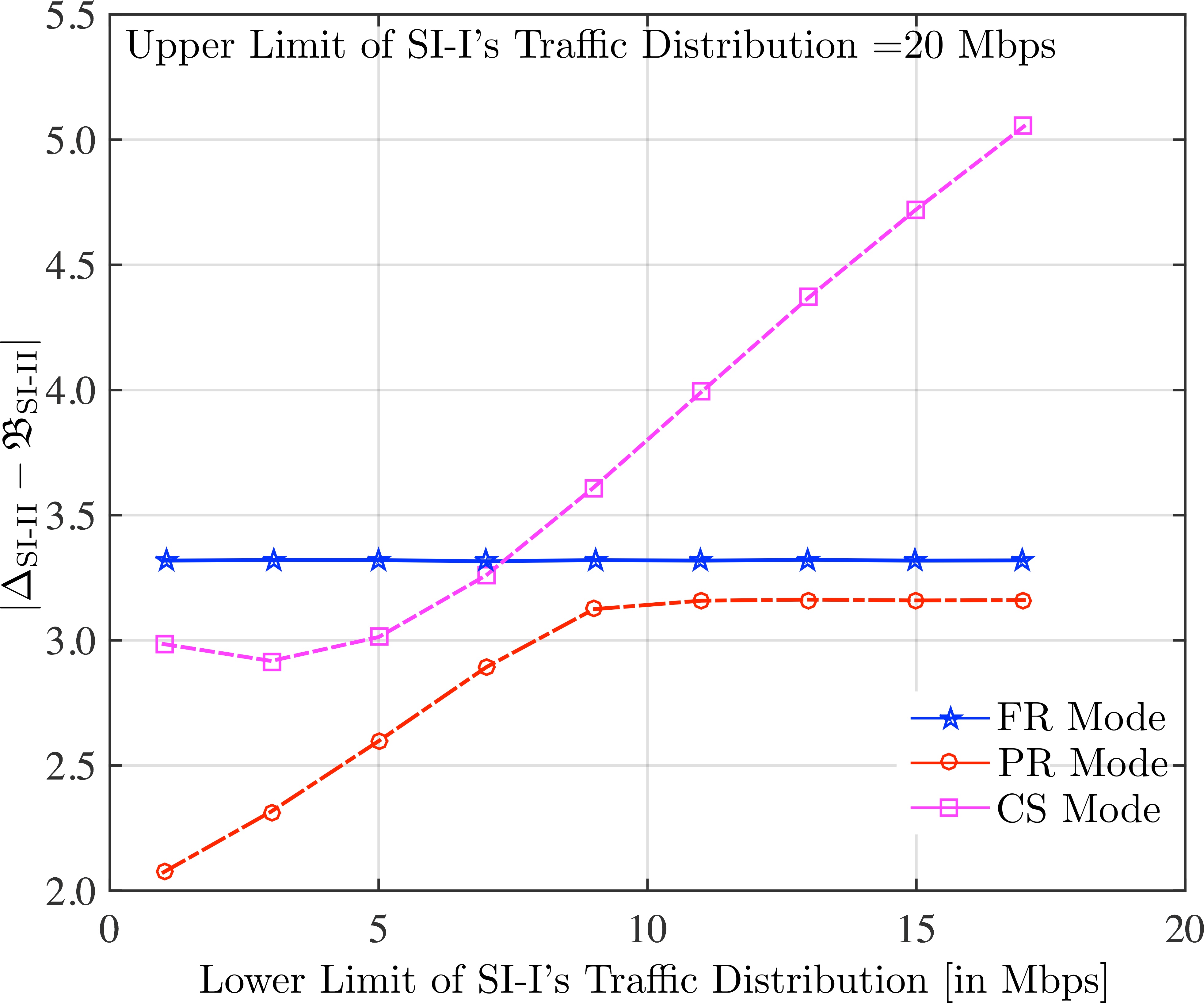}
\caption{Deviation characteristics for SI-II with different modes of spectrum coordination policies when the lower limit of the traffic distribution under SI-I varies. Here, $\xi_{\text{SI-I}}=\xi_{\text{SI-II}}=0.5$ and $\Theta=0$ MHz.}
\label{perffig2}
\end{figure}
In Fig.~\ref{perffig2}, we again investigate the characteristics of the deviation measure $|\Delta_{p}-\mathfrak{B}_p|$, however, this time it is for SI-II, for all the considered spectrum coordination policies when lower limit of the SI's spectrum demand distribution varies with the upper limit being fixed to 20 MHz. Note that when the lower limit of the SI-I's spectrum demand distribution increases, -i.e., the likelihood of SI-I demanding more spectrum also increases. Again, for the PR mode, the SIs' principal shares are taken as $\bar{\xi}_{\text{SI-I}}=\bar{\xi}_{\text{SI-I}}=0.5$. We observe similar deviation characteristics as in Fig.~\ref{perffig1} with PR mode providing the best deviation performance for SI-II. Note that if the demands from SIs can always be satisfied with the available spectrum and/or when the NH-$\mu$O has sufficient bandwidth, choosing any spectrum coordination policy would perform desired job. However, as long as the spectrum utilization efficiency is concerned, PR mode seems to provide the best performance. Also note that based on the definition of the deviation measure employed, the spectrum allocation policies incur lower deviation for all the SIs when their spectrum demands follow and/or are consistent with their principal shares.

\begin{figure}
\centering
\includegraphics[scale=.08]{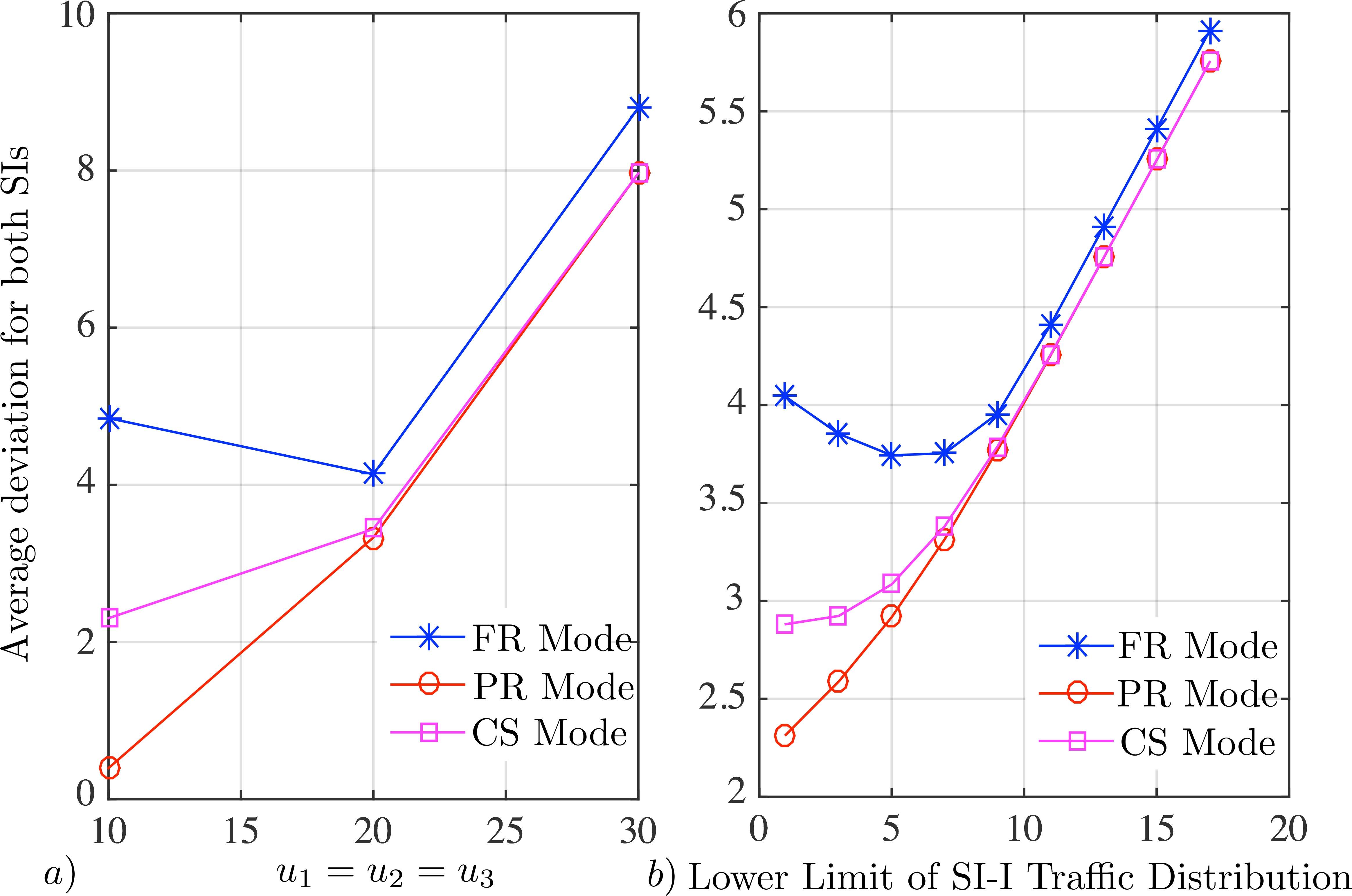}
\caption{Average deviation characteristics for both the SIs with different modes of spectrum coordination policies when a) the number of users belonging to the MNOs under SI-II varies and b) the lower limit of the traffic distribution under SI-I varies. Here, $\xi_{\text{SI-I}}=\xi_{\text{SI-II}}=0.5$ and $\Theta=0$ MHz.}
\label{perffig5}
\end{figure}
We noticed that the deviation characteristics of SI-II for the settings in Fig.~\ref{perffig1} and the deviation characteristics of SI-I for the settings in Fig.~\ref{perffig2} are not similar to that of SI-I and SI-II in Fig.~\ref{perffig1} and Fig.~\ref{perffig2}, respectively. When the traffic in both the scenarios is higher than some values dues to higher number of users in SI-II or higher lower limit of traffic distribution of SI-I, the CS mode performs better than the PR mode. Even for FR mode, the curves follow the nature of CS mode in Fig.~\ref{perffig2}.
In order to find the best spectrum coordination policy suited for both the SIs under the settings of  Fig.~\ref{perffig1} and Fig.~\ref{perffig2}, we calculated the average devision and provide the deviation characteristics in Fig.~\ref{perffig5}a and Fig.~\ref{perffig5}b, respectively. We find that the PR mode still performs better compared to the other spectrum coordination policies.

\begin{figure}
\centering
\includegraphics[scale=.08]{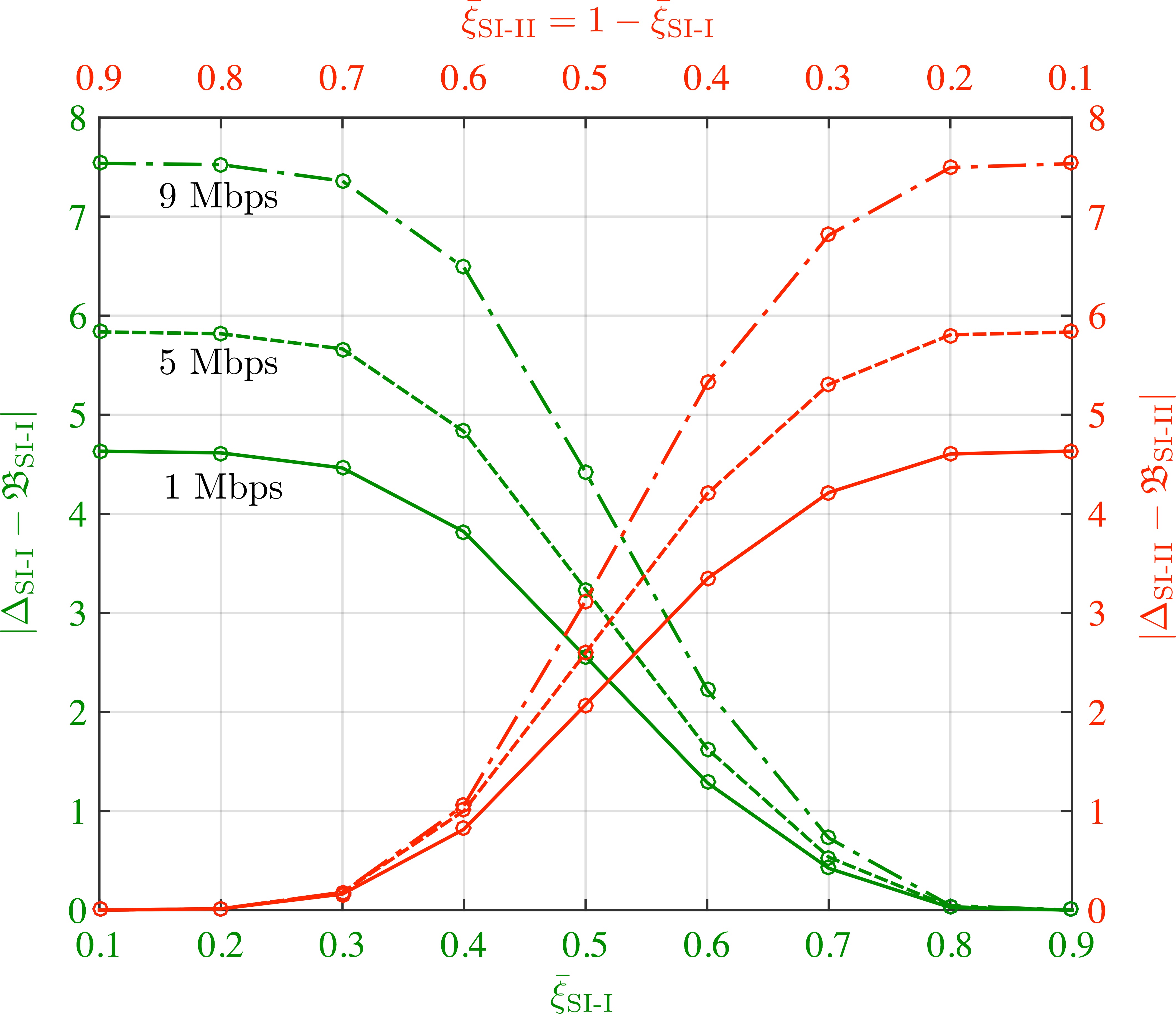}
\caption{Deviation characteristics for both the SIs with PR mode spectrum coordination policy when the the principal shares of the SIs vary. The number of users belonging to the MNOs under SI-II is fixed to 30, -i.e., $u_i=u_2=u_3=30,$ while the lower limit of SI-I's uniform traffic distribution is varied. The traffic demand for $\mu$O own ultra-reliable business operation is set to 5 Mbps.}
\label{perffig3}
\end{figure}
In Fig.~\ref{perffig3}, we analyze the characteristics of the deviation parameter when the SIs have varying principal shares with  $\bar{\xi}_{\text{SI-I}}+\bar{\xi}_{\text{SI-II}}=1$. The number of users belonging to the MNOs under SI-II is fixed to 30, -i.e., $u_i=u_2=u_3=30,$ while the lower limit of SI-I's uniform traffic distribution is varied. The traffic demand for $\mu$O own ultra-reliable business operation is set to 5 Mbps. The plot provides an answer to the question on how much of principal share one SI needs to have in order to satisfy some threshold of deviation parameter. We can clearly observe that the SIs encounter different levels of deviation depending on its amount of principal share. It is obvious that the higher the principal share, the lower is the deviation.
However, having higher principal share in the shareable spectrum will cost the SI, especially the MNOs under SI-II with higher investment or capital. Therefore, the SI and/or the MNOs will prefer to obtain the lowest possible amount of principal share that satisfies its deviation threshold. An SI with higher principal share will experience lower deviation compared to the one with lower principal share if their spectrum demands follow the same distribution. Since SI-I has a different traffic distribution than SI-II, we notice different deviation characteristics for the SIs. The SI-II has better deviation performance (lower deviation) compared to SI-I for the same principal share.

\begin{figure}
\centering
\includegraphics[scale=.08]{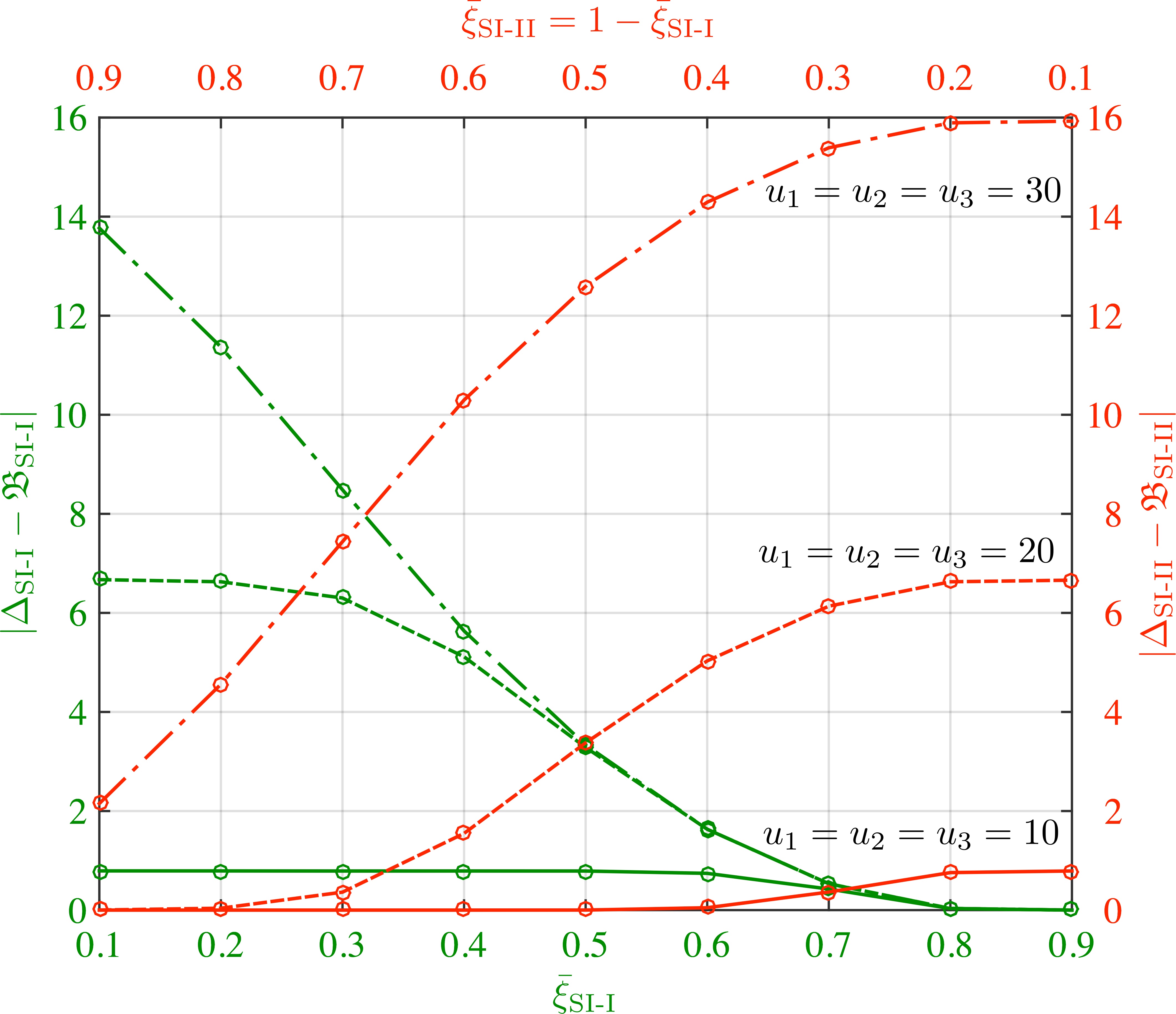}
\caption{Deviation characteristics for both the SIs with PR mode spectrum coordination policy when the the principal shares of the SIs vary. The lower limit of SI-I's uniform traffic distribution is fixed to 1 Mbps while the number of users belonging to the MNOs under SI-II is varied. The traffic demand for $\mu$O's own ultra-reliable business operation is set to 5 Mbps.}
\label{perffig4}
\end{figure}
Similar to Fig.~\ref{perffig3}, in Fig.~\ref{perffig4}, we analyze the characteristics of the deviation parameter when the SIs have varying principal shares. However, in Fig.~\ref{perffig4}, the lower limit of SI-I's uniform traffic distribution is fixed to 1 Mbps while the number of users belonging to the MNOs under SI-II is varied as shown. We notice that when the MNOs user SI-II have small number of users to serve, -i.e., comparatively lower spectrum demand from SI-II, both the SIs experiences very good deviation performance as SI-II will, most of the times, favor SI-I to have access to its additional spectrum. However, like in Fig.~\ref{perffig3}, SI-II still requires lower amount of principal share to satisfy its desired deviation threshold. As the number of users per MNO increases to 20, the deviation curves for SI-I and SI-II produces almost symmetry-based reciprocity. A slight increase in principal share of one SI has a great impact on the deviation measure of the other SI. We can see that the values of the deviation parameters for both the SIs increase compared to the case where the MNOs under SI-II served small number of users. Therefore, instantaneous traffic variation in one SIs has strong impact on the performance of the other SI.

\section{Conclusion}

In this paper, we conceive an advanced wireless networks deployment framework that facilitates $\mu$Os to become service providers.
Our proposed framework serve two-fold advantages, such as it gives the venue owner its own managed wireless networks tailored to its very specific requirements, and it also brings out cost savings and coverage extension for MNOs and efficiency of resources that arise from sharing wireless networks, and delivering the network capacity into high density venues. 
The NH-$\mu$O offers heterogeneous use-case and/or service aggregation. It addresses the provisions of providing localized and/or local context based services by the venue owners, and MNOs' need to have coverage extension and/or offload their subscribers traffic to more scalable indoor wireless networked system. This paper mainly deals with the deployment issues in next generation networks.
We design algorithms for dynamic resource sharing across SIs tracking network load dynamics, which is challenging as it involves inter-SI as well as multi-MNO sharing policies.

\section*{Acknowledgement}
This research was conducted under a contract of R\&D for Expansion of Radio Wave Resources, organized by the Ministry of Internal Affairs and Communications, Japan.


\end{document}